\documentclass[lettersize,journal]{IEEEtran}
\usepackage[colorlinks,linkcolor=magenta,anchorcolor=magenta,citecolor=magenta,bookmarks=true]{hyperref}  
\usepackage[T1]{fontenc}
\ifCLASSINFOpdf
\usepackage[pdftex]{graphicx} 
\DeclareGraphicsExtensions{.eps,.pdf,.jpeg,.png}
\else
\usepackage[dvips]{graphicx}
\fi
\usepackage[compress,nospace]{cite}
\usepackage[cmex10]{amsmath}
\usepackage{setspace}
\usepackage{enumitem}
\usepackage[ruled]{algorithm2e}

\usepackage{epstopdf,amsthm,stfloats,siunitx,amssymb,wasysym,array,url, color,subfigure} 

\usepackage{footnote}
\makesavenoteenv{tabular}
\usepackage{soul}

\begin{document}
	%
	\title{RIS-ADMM: A RIS and ADMM-Based Passive and Sparse Sensing Method With Interference Removal}
	%
	%
	%
	
	\author{Peng~Chen,~\IEEEmembership{Senior Member,~IEEE,} 
		Zhimin~Chen,~\IEEEmembership{Member,~IEEE,}
		Pu~Miao,~\IEEEmembership{Member,~IEEE,}
		Yun~Chen,~\IEEEmembership{Member,~IEEE} 
		
		\thanks{This work was supported in part by the Natural Science Foundation for Excellent Young Scholars of Jiangsu Province  under Grant BK20220128, the Natural Science Foundation of Shanghai under Grant 22ZR1425200, the National Key R\&D Program of China under Grant 2019YFE0120700, the National Key Laboratory on Electromagnetic Environmental Effects and Electro-optical Engineering under Grant JCKYS2023LD6, the Open Fund of State Key Laboratory of Integrated Chips and Systems, and the National Natural Science Foundation of China under Grant  61801112. \textit{(Corresponding author: Zhimin Chen)}}	
		 
		\thanks{P.~Chen is with the State Key Laboratory of Millimeter Waves, Southeast University, Nanjing 210096, China, and also with the State Key Laboratory of Integrated Chips and Systems, Fudan University, Shanghai 201203, China (e-mail: chenpengseu@seu.edu.cn).}
		\thanks{Z.~Chen is with the School of Electronic and Information, Shanghai Dianji University, Shanghai 201306, China (e-mail: chenzm@sdju.edu.cn).} 
		\thanks{P.~Miao is with the School of Electronic and Information Engineering, Qingdao University, Shandong 266071, China (e-mail: mpvae@qdu.edu.cn).} 
		\thanks{Y.~Chen is with the State Key Laboratory of ASIC and System and the Microelectronics School, Fudan University, Shanghai 201203, China (e-mail: chenyun@fudan.edu.cn).} 
	}

	\markboth{IEEE Communications letters}%
	{Shell \MakeLowercase{\textit{et al.}}: Bare Demo of IEEEtran.cls for IEEE Journals}
	
	\maketitle
	
	\begin{abstract}
		Reconfigurable Intelligent Surfaces (RIS) emerge as promising technologies in future radar and wireless communication domains. This letter addresses the passive sensing issue utilizing wireless communication signals and RIS amidst interference from wireless access points (APs). We introduce an atomic norm minimization (ANM) approach to leverage spatial domain target sparsity and estimate the direction of arrival (DOA). However, the conventional semidefinite programming (SDP)-based solutions for the ANM problem are complex and lack efficient realization. Consequently, we propose a RIS-ADMM method, an innovative alternating direction method of multipliers (ADMM)-based iterative approach. This method yields closed-form expressions and effectively suppresses interference signals. Simulation outcomes affirm that our RIS-ADMM method surpasses existing techniques in DOA estimation accuracy while maintaining low computational complexity. The code for the proposed method is available online \url{https://github.com/chenpengseu/RIS-ADMM.git}.
	\end{abstract}
	

	\section{Introduction}
	\IEEEPARstart{R}{ecently}, the integrated sensing and communication (ISAC) system has garnered increasing attention for its capability to perform both wireless communication and target localization within a single apparatus~\cite{9729741}. The tradeoff between the detection performance and the communication rate in the ISAC system is give in~\cite{10124135}. The ISAC system operates in two modes: active and passive. The active mode transmits additional signals for sensing, whereas the passive mode utilizes only the communication signals for this purpose. The latter, when employed for localizing an unregistered target using communication signals, is often referred to as device-free sensing~\cite{9724258}. Notably, passive sensing systems are advantageous for their convenient implementation and compatibility with existing communication infrastructures~\cite{9198891}.
	
	Target localization in such systems can be achieved through the analysis of signal delays or directions. Delay estimation typically requires a synchronization reference signal from the base station (BS) or wireless access point (AP), necessitating an additional synchronization link. Conversely, direction estimation does not require this synchronization signal and is thus more convenient for system integration. However, traditional direction-finding systems often involve multiple receiving channels, leading to increased cost and complexity. Recently, the advent of reconfigurable intelligent surfaces (RIS) has been proposed as a solution to enhance communication and radar performance, as well as to facilitate direction of arrival (DOA) estimation with a single, fully functional receiving channel by modulating the reflected signal amplitudes or phases~\cite{8811733,8910627}. 
	
	There have been several propositions for RIS-based methods in communication and sensing systems. For instance, a RIS-enhanced spectrum sensing system is proposed in~\cite{9761384}, where the detection probability is assessed to determine the required number of elements for high detection accuracy. In~\cite{9593143}, RIS is employed to achieve both radar and communication functionalities in a mmWave system, with a specially designed codebook to localize targets while maintaining robust communication. In~\cite{10283788}, an expectation-maximization (EM) -based Bayesian learning algorithm is proposed to estimate the mmWave channel by exploring the angular sparsity in the RIS-adided ISAC system. Ref.~\cite{9591331} optimizes the joint waveform and RIS phases to reduce multi-user interference and ensure DOA estimation accuracy. RIS's role extends to location-aware communication in beyond 5G networks as discussed in~\cite{9729782,9215972}. The discretized effects of phase shifts due to hardware constraints are analyzed in~\cite{9223720}. Additionally, the issues surrounding target detection in radar systems using RIS are explored in~\cite{9454375,9725255}. Furthermore, stacked intelligent metasurfaces (SIM) as a revolutionary technology of the RIS outperform the single-layer RIS by stacking multiple metasurface layers, can be also used to get the DOA information by controlling the reflected signals~\cite{10158690,10379500}.

	
In this letter, we explore the passive sensing challenges in scenarios impacted by interference signals from AP using RIS. We introduce an optimized solution based on atomic norm minimization (ANM) to address these challenges effectively. Traditional methodologies predominantly utilize semidefinite programming (SDP)-based strategies for solving ANM problems; however, these approaches are often constrained by their substantial computational complexity and inefficiency. To circumvent these limitations, we introduce a cutting-edge RIS-ADMM algorithm. This algorithm, grounded in the alternating direction method of multipliers (ADMM), is not only adept at resolving the ANM problem but also integrates an effective interference mitigation mechanism. Moreover, we have derived explicit closed-form expressions for the iterations within the RIS-ADMM algorithm, thereby significantly enhancing its practical applicability. The paper concludes with a thorough comparative analysis of the proposed method against existing techniques, with a specific focus on the accuracy of DOA estimation and computational efficiency.
	
	
	The structure of this paper is methodically organized as follows. Section~\ref{sec2} details the formulation of the RIS-based sensing system model. Section~\ref{sec3} introduces the proposed RIS-ADMM method. Section~\ref{sec4} presents a comprehensive analysis of the simulation results. Finally, Section~\ref{sec5} provides the conclusions.
	

	\section{The RIS-Based Sensing System Model}\label{sec2}
	
	\begin{figure}
		\centering
		\includegraphics[width=2.2in]{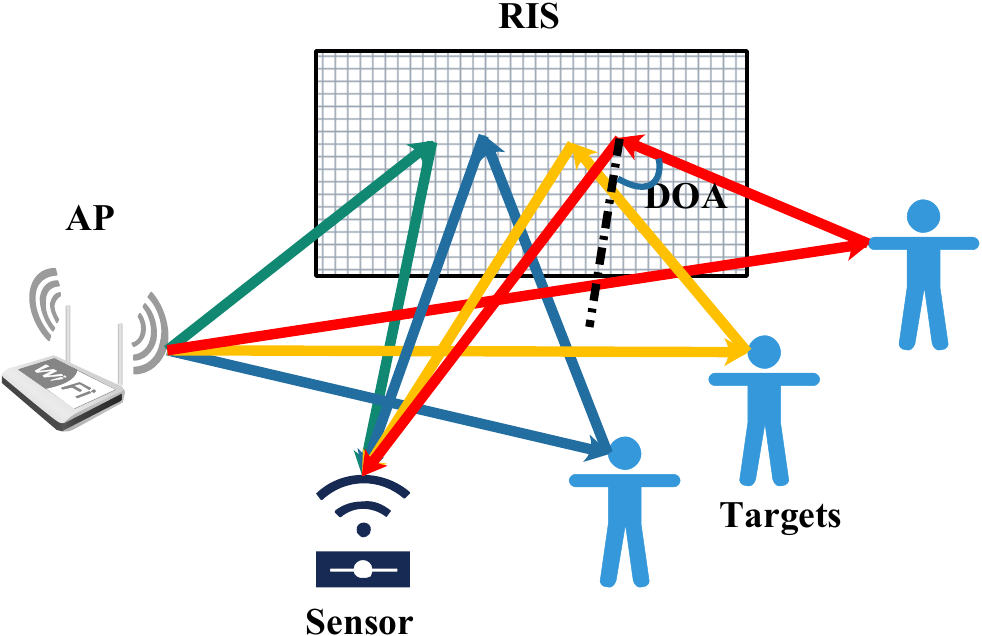}
		\caption{The system model of the passive sensing using RIS.}
		\label{sys}
	\end{figure}

	In this work, we investigate a RIS-based sensing system as depicted in Fig.~\ref{sys}. An AP facilitates wireless communication, with its signals serving the dual purpose of passive sensing. The RIS reflects these communication signals, which are subsequently received by a sensor equipped with a single receiving channel. Our objective is to estimate the DOA between the targets and the RIS by modulating the amplitude and phases of the RIS elements across various time slots. Note that the path between the AP and the sensor, and the path between the target and the sensor are ignored, since the electromagnetic waves at these paths cannot be controlled by the RIS. Additionally, signals subjected to multiple reflections are deliberately omitted due to their inherently low power levels.
	
	Under the far-field assumption, the received signal $\boldsymbol{r}\in\mathbb{C}^{N\times 1}$ at the sensor over $N$ time slots and with $K$ targets is articulated as
	\begin{align}\label{eq1}
		\boldsymbol{r} = \boldsymbol{GA}(\boldsymbol{\theta})\boldsymbol{s}+\boldsymbol{Ga}(\psi)q+\boldsymbol{w},
	\end{align}
	where $\boldsymbol{s}\in\mathbb{C}^{K\times 1}$ represents the reflected signal and $\boldsymbol{w}\in\mathbb{C}^{N\times 1}$ denotes the additive white Gaussian noise (AWGN) with a variance of $\sigma_{\text{w}}^{2}$. The measurement matrix $\boldsymbol{G}\in\mathbb{C}^{N\times M}$ encompasses $N$ measurements and $M$ RIS elements. This measurement matrix can improve the signal-to-noise ratio (SNR) of the reflected signals by optimizing the reflected beam~\cite{9687593}. Notably, an interference signal $q\in\mathbb{C}$ emanates from the direct path between the AP and the RIS, affecting the sensing system. The DOA for the $k$-th target (where $k=0,1,\dots,K-1$) relative to the RIS is denoted as $\theta_k$, and the DOA between the AP and RIS is represented as $\psi$. The steering matrix is defined as $\boldsymbol{A}(\boldsymbol{\theta})\triangleq\begin{bmatrix}
		\boldsymbol{a}(\theta_0),\boldsymbol{a}(\theta_1),\dots, \boldsymbol{a}(\theta_{K-1})
	\end{bmatrix}\in\mathbb{C}^{M\times K}$, with the steering vector $\boldsymbol{a}(\theta)$ expressed as $\boldsymbol{a}(\theta)\triangleq\begin{bmatrix}
		1,e^{j2\pi d \sin\theta},\dots,e^{j2\pi (M-1) d\sin\theta}
	\end{bmatrix}^{\text{T}}\in\mathbb{C}^{M\times 1}$, where $j\triangleq\sqrt{-1}$, and $d$ signifies the normalized distance between adjacent elements relative to the wavelength. The system model  addressed in this paper can be effectively adapted for application in a three-dimensional (3D) model by appropriately modifying the steering vector. In this paper, we delve into the passive sensing issue, wherein both the reflected signals $\boldsymbol{s}$ and the interference signal $q$ are unknown entities. Employing the system model delineated in (\ref{eq1}), our objective is to deduce the DOA $\boldsymbol{\theta}$ from the received signal $\boldsymbol{r}$, amidst the presence of an extraneous interference signal.
	
	\section{RIS-ADMM Method for the DOA Estimation With Interference Removal}\label{sec3}
	In this study, we focus on leveraging target sparsity to enhance the DOA estimation performance within a passive sensing framework. Furthermore, to address the \emph{off-grid} issues inherent in compressed sensing (CS) methods utilizing a dictionary matrix, we propose formulating an ANM problem within our RIS-ADMM methodology. Initially, considering the presence of an interference signal, the DOA estimation issue is redefined as a sparse reconstruction problem as follows
	\begin{align}
		\min_{\boldsymbol{x},q}\, &\|\boldsymbol{r}-\boldsymbol{Gx}-\boldsymbol{c}q\|^2_2+\rho \|\boldsymbol{x}\|_{\mathcal{A}},\label{eq4}
	\end{align}
	where $\boldsymbol{c}\triangleq\boldsymbol{Ga}(\psi)$ and $\|\cdot\|_{\mathcal{A}}$ denotes the atomic norm. The parameter $\rho$ regulates the sparsity of the reconstructed signal $\boldsymbol{x}$ and is typically set as $\rho \approx \sigma_{\text{w}}\sqrt{M\log M}$. The atomic norm is defined as
	$
	\|\boldsymbol{x}\|_{\mathcal{A}}=\inf_{\boldsymbol{u},t}\Bigg\{ 
	\frac{\operatorname{Tr}\{\operatorname{Toep}(\boldsymbol{u})\}}{2M}+\frac{t}{2}: \begin{pmatrix}
		\operatorname{Toep}(\boldsymbol{u})&\boldsymbol{x}\\
		\boldsymbol{x}^{\text{H}} & t
	\end{pmatrix}\succeq 0
	\Bigg\}$~\cite{9398559}. 
	
	Subsequently, we reformulate the ANM problem as
	\begin{align}\label{eq6}
		\min_{\boldsymbol{x},q,\boldsymbol{u},t}\, &\|\boldsymbol{r}-\boldsymbol{Gx}-\boldsymbol{c}q\|^2_2+\frac{\rho}{2M}\operatorname{Tr}\{\operatorname{Toep}(\boldsymbol{u})\}+\frac{\rho}{2}t.\\
		\text{s.t.}\, & \begin{pmatrix}
			\operatorname{Toep}(\boldsymbol{u})&\boldsymbol{x}\\
			\boldsymbol{x}^{\text{H}} & t
		\end{pmatrix}\succeq 0\notag.
	\end{align}
	While the passive sensing system using RIS can reconstruct the sparse signal $\boldsymbol{x}$ by solving the SDP-based method (\ref{eq6}), the practical application involving a large number of RIS elements and low-cost components challenges the direct application of this method due to its high computational complexity. Therefore, we propose an ADMM-based method to efficiently tackle the SDP problem.
	
	The optimization problem is subsequently expressed as
	\begin{align}
		\min_{\boldsymbol{x},q,\boldsymbol{u},t,\boldsymbol{z},\boldsymbol{Y}}\, &\|\boldsymbol{r}-\boldsymbol{z}\|^2_2+\frac{\rho}{2}(u_0+t).\\
		\text{s.t.}\, & \boldsymbol{Y}=\begin{pmatrix}
			\operatorname{Toep}(\boldsymbol{u})&\boldsymbol{x}\\
			\boldsymbol{x}^{\text{H}} & t
		\end{pmatrix},\, \boldsymbol{Y}\succeq 0,\, \boldsymbol{z} = \boldsymbol{Gx}+\boldsymbol{c}q.\notag
	\end{align}
	Then, by introducing parameters $\boldsymbol{P}$ and $\boldsymbol{w}$, the corresponding augmented Lagrangian is
	\begin{align}
		&\mathcal{L}_{\tau}(\boldsymbol{x},q,\boldsymbol{u},t,\boldsymbol{z},\boldsymbol{Y},\boldsymbol{P},\boldsymbol{w}) = \|\boldsymbol{r}-\boldsymbol{z}\|^2_2+\frac{\rho}{2}(u_0+t)\notag\\
		&\quad +\Bigg\langle \boldsymbol{P},\boldsymbol{Y}-\begin{pmatrix}
			\operatorname{Toep}(\boldsymbol{u})&\boldsymbol{x}\\
			\boldsymbol{x}^{\text{H}} & t
		\end{pmatrix}\Bigg\rangle+\langle \boldsymbol{Gx}+\boldsymbol{c}q-\boldsymbol{z},\boldsymbol{w}\rangle\notag\\
		&\quad + \tau \left\|\boldsymbol{Y}-\begin{pmatrix}
			\operatorname{Toep}(\boldsymbol{u})&\boldsymbol{x}\\
			\boldsymbol{x}^{\text{H}} & t
		\end{pmatrix}\right\|^2_{\text{F}}+\tau\|\boldsymbol{Gx}+\boldsymbol{c}q-\boldsymbol{z}\|^2_2,
	\end{align}
	where we denote the vector inner product operation as $\langle \boldsymbol{x},\boldsymbol{y} \rangle\triangleq \mathcal{R}\{\boldsymbol{y}^{\text{H}}\boldsymbol{x}\}$ and the matrix inner product operation as $\langle \boldsymbol{X},\boldsymbol{Y} \rangle\triangleq \mathcal{R}\{\operatorname{Tr}(\boldsymbol{Y}^{\text{H}}\boldsymbol{X})\}$. The parameter $\tau$ is for the robustness of the reconstruction signal, and can be chosen as $\tau\approx 0.5   \sigma_{\text{w}}\sqrt{M\log M}$.
	
		\begin{algorithm} 
		\caption{RIS-ADMM} \label{alg1}
		\LinesNumbered
		\setstretch{0.9}
		\KwIn{The received signal $\boldsymbol{r}$, the number iteration $i_{\text{num}}$, and the number of measurements $M$.} 
		\textbf{Initialization:} $q=0$, $\boldsymbol{x}=\boldsymbol{0}$, $\boldsymbol{u}=\boldsymbol{0}$, $t=0$,$\boldsymbol{z}=\boldsymbol{0}$, $\boldsymbol{P}=\boldsymbol{0}$, $\boldsymbol{Y}=\boldsymbol{0}$, $\boldsymbol{P}\triangleq\begin{pmatrix}
			\boldsymbol{P}_1 & \boldsymbol{p}_2\\
			\boldsymbol{p}_2^{\text{H}} & p_3
		\end{pmatrix}$,  $\boldsymbol{Y}\triangleq\begin{pmatrix}
			\boldsymbol{Y}_1 & \boldsymbol{y}_2\\
			\boldsymbol{y}_2^{\text{H}} & y_3
		\end{pmatrix}$, $\boldsymbol{w}=\boldsymbol{0}$ , $\boldsymbol{c}' = \frac{1}{\tau \|\boldsymbol{c}\|^2_2}\boldsymbol{c}$, $\boldsymbol{G}'=(\tau\boldsymbol{G}^{\text{H}}\boldsymbol{G}+2\tau\boldsymbol{I})^{-1}$, $i=0$\;
		\While{$i<i_{\text{num}}$}
		{
			$q = \boldsymbol{c}'^{\text{H}}(\tau \boldsymbol{z}-0.5\boldsymbol{w}-\tau \boldsymbol{Gx})$\;
			$\boldsymbol{x}=\boldsymbol{G}'\big[2\tau\boldsymbol{y}_2+\boldsymbol{p}_2 +\boldsymbol{G}^{\text{H}}(\tau \boldsymbol{z}-0.5\boldsymbol{w}-\tau q \boldsymbol{c})\big]$,$m=0$\;
			\While{$m<M$} {
				\eIf{$m=0$} {$ u_m =\frac{ \operatorname{Tr}(\boldsymbol{P}_1,m) -\rho }{2\tau M}$\;
				}{
					$ u_m =\frac{ \operatorname{Tr}(\boldsymbol{P}_1,m)+2\tau\operatorname{Tr}(\boldsymbol{Y}_1,m) }{2\tau (M-m)}$\;
				}
				$m\leftarrow m+1$\;
			} 
			$t=y_3+\frac{p_3}{2\tau}-\frac{\rho}{4\tau}$, 
			$\boldsymbol{Q}\leftarrow\begin{pmatrix}
				\operatorname{Toep}(\boldsymbol{u})&\boldsymbol{x}\\
				\boldsymbol{x}^{\text{H}} & t
			\end{pmatrix}$\; 
			$ 
			\boldsymbol{z}=\frac{1}{1+\tau}\left[\boldsymbol{r}+0.5\boldsymbol{w}+\tau(\boldsymbol{Gx}+\boldsymbol{c}q)\right]$\;
			Use the eigenvalue decomposition $
			\boldsymbol{Q}-\frac{1}{2\tau}\boldsymbol{P}=\boldsymbol{U\Lambda U}^{\text{H}}$. $
			\boldsymbol{Y}=  \boldsymbol{U}\boldsymbol{\Lambda}'\boldsymbol{U}^{\text{H}}$,
			where the diagonal matrix $\boldsymbol{\Lambda}'$ is obtained by setting the negative entries of being $0$\;
			$
			\boldsymbol{P}\leftarrow \boldsymbol{P}+\rho\left(\boldsymbol{Y}-\boldsymbol{Q}\right)$\;
			$
			\boldsymbol{w}\leftarrow \boldsymbol{w}+0.5\rho(\boldsymbol{Gx}+\boldsymbol{c}q-\boldsymbol{z})$,
			$i\leftarrow i+1$\;
		}
		\KwOut{$\boldsymbol{x}$.} 
	\end{algorithm}
	
	 In the ADMM method~\cite{boyd2011distributed}, by denoting $(\cdot)^i$ as the value of the parameter at the $i$-th iteration, all the unknown parameters during the $i$-th iteration can be obtained as follows:
	\begin{itemize}[topsep=-4pt]
		\item For the unknown parameters $\boldsymbol{x}^{i+1}$, $q^{i+1}$,$\boldsymbol{u}^{i+1}$,$t^{i+1}$,$\boldsymbol{z}^{i+1}$, we have
		\begin{align}
			& \{\boldsymbol{x}^{i+1},q^{i+1},\boldsymbol{u}^{i+1},t^{i+1},\boldsymbol{z}^{i+1}\}\\
			&\qquad \qquad=\arg\min_{\boldsymbol{x},q,\boldsymbol{u},t,\boldsymbol{z}} \mathcal{L}_{\tau}(\boldsymbol{x},q,\boldsymbol{u},t,\boldsymbol{z},\boldsymbol{Y}^{i},\boldsymbol{P}^{i},\boldsymbol{w}^{i}).  \notag  
		\end{align}
		\item For the semidefinite matrix $\boldsymbol{Y}$, it can be updated as
		\begin{align}
			&\boldsymbol{Y}^{i+1} = \arg\min_{\boldsymbol{Y}\succeq 0} \mathcal{L}_{\tau}(\boldsymbol{x}^{i+1},q^{i+1},\boldsymbol{u}^{i+1},t^{i+1},\boldsymbol{z}^{i+1},\notag\\
			&\qquad\qquad\qquad \boldsymbol{Y},\boldsymbol{P}^{i},\boldsymbol{w}^{i}).
		\end{align}
		\item Update the parameters $\boldsymbol{P}$ and $\boldsymbol{w}$ as
		\begin{align}
			&\boldsymbol{P}^{i+1}=\boldsymbol{P}^{i}+\rho\nabla_{\boldsymbol{P}^*}\mathcal{L}_{\tau}(\boldsymbol{x}^{i+1},q^{i+1},\boldsymbol{u}^{i+1},\notag\\
			&\qquad\qquad\qquad t^{i+1},\boldsymbol{z}^{i+1},\boldsymbol{Y}^{i+1},\boldsymbol{P},\boldsymbol{w}^{i}),\\
			&\boldsymbol{w}^{i+1}=\boldsymbol{w}^{i} +\rho\nabla_{\boldsymbol{w}^*}\mathcal{L}_{\tau}(\boldsymbol{x}^{i+1},q^{i+1},\boldsymbol{u}^{i+1},\notag\\
			&\qquad\qquad\qquad t^{i+1},\boldsymbol{z}^{i+1},\boldsymbol{Y}^{i+1},\boldsymbol{P}^{i+1},\boldsymbol{w}).
		\end{align}
	\end{itemize}
	In conclusion, the RIS-ADMM method presents a formidable alternative to traditional methods, providing a closed-form solution to the iterative process. Detailed derivations are presented in Appendix~\ref{ap1}. The proposed method employs an ADMM-based approach to address a type of the least absolute shrinkage and selection operator (LASSO) problem. Consistent with findings in~\cite{10292437,7423789}, it is demonstrated that the ADMM-based methodology is capable of converging to a solution that fulfills the Karush-Kuhn-Tucker (KKT) conditions. This convergence property not only underscores the efficacy of the proposed method in solving the optimization problem,  but also reinforces its applicability in scenarios where compliance with the KKT condition is crucial for obtaining optimal solution.
	
	While we successfully reconstruct the sparse signal $\boldsymbol{x}$, the actual DOAs remain to be determined. For this, we employ the multiple signal classification (MUSIC) method with a single snapshot, such as the Henkel-based MUSIC method~\cite{liao2016music}, to estimate the spatial spectrum and subsequently the DOAs by locating the peak positions. Notably, the computational complexity of the RIS-ADMM method, dominated by eigenvalue decomposition, is $\mathcal{O}((M+1)^3)$, offering a significant reduction from the $\mathcal{O}((M+1)^4)$ complexity of the SDP-based methods for problem (\ref{eq6}).

	%


%

		\begin{figure}
		\centering
		\includegraphics[width=2in]{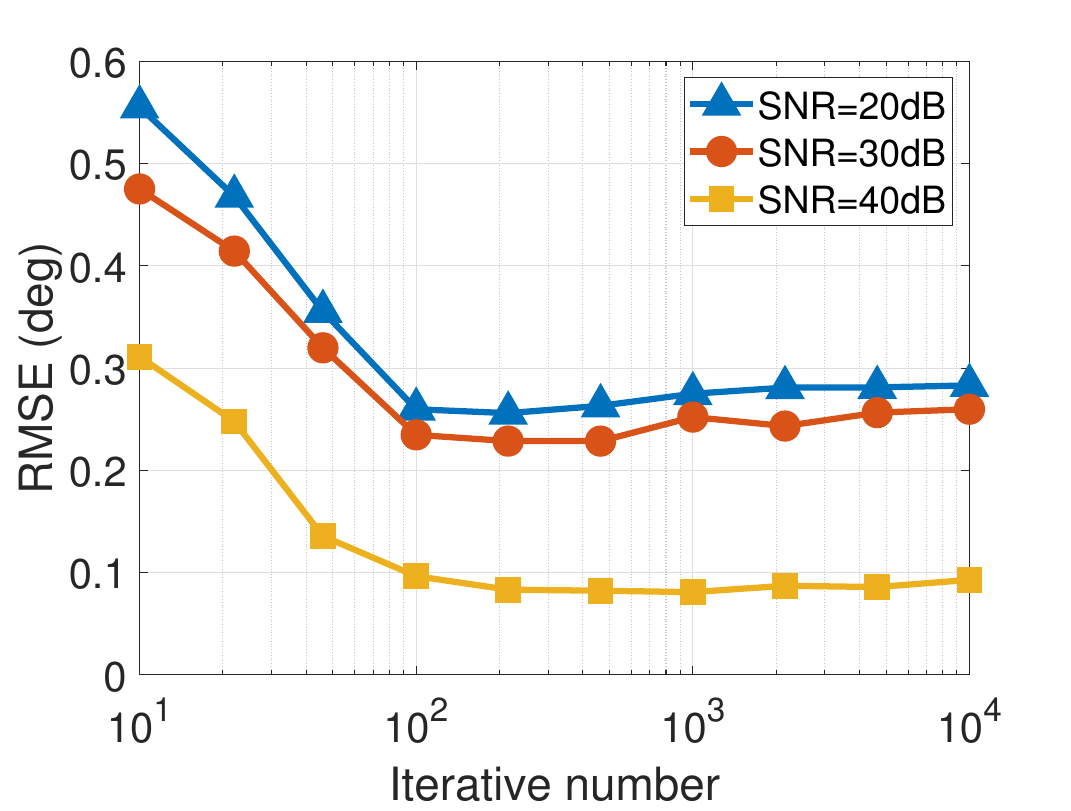}
		\caption{The estimated performance with different iterative numbers.}
		\label{iternum}
	\end{figure}
	
	\section{Simulation Results}\label{sec4}
	In this section, we demonstrate the efficacy of the proposed RIS-ADMM method through extensive simulations. The source code for the implemented method is publicly accessible at \url{https://github.com/chenpengseu/RIS-ADMM.git}. The simulation parameters are set as follows: the number of RIS elements is $M=64$; the number of time slots for measurements is $N=32$; the number of targets is $K=3$; the normalized element spacing is $d=0.5$; and the spatial angle spans from $-\ang{90}$ to $\ang{90}$. 
	
	Initially, we explore the estimation performance of the RIS-ADMM method over various iteration counts, as depicted in Fig.~\ref{iternum}. Here, the performance is quantified by the Root Mean Square Error (RMSE) defined as $
	\text{RMSE}=\sqrt{1/(K N_{\text{mc}})\sum_{i_{\text{mc}}=0}^{N_{\text{mc}}}\|\hat{\boldsymbol{\theta}}_{i_{\text{mc}}}-\boldsymbol{\theta}_{i_{\text{mc}}}\|^2_2}$,
	where $N_{\text{mc}}$ represents the number of Monte Carlo trials, $\hat{\boldsymbol{\theta}}_{i_{\text{mc}}}$ is the estimated DOA, and $\boldsymbol{\theta}_{i_{\text{mc}}}$ is the actual DOA. The results suggest that an increase in iterations leads to improved performance, stabilizing beyond $10^2$ iterations. Consequently, for subsequent simulations, we fix the number of iterations at $10^2$. 
	
\begin{figure}
	\centering
	\includegraphics[width=2.8in]{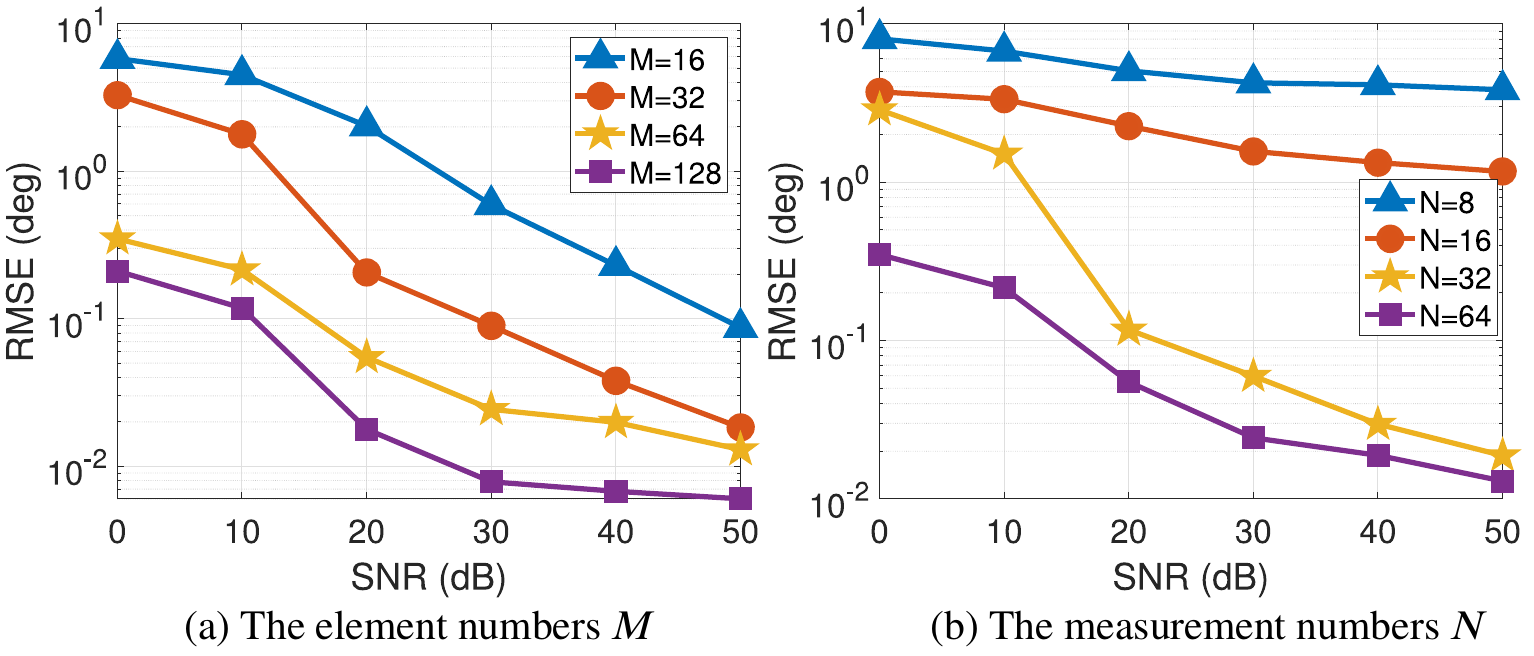}
	\caption{The estimated performance with different system parameters.}
	\label{MN}
\end{figure}

	Furthermore, we investigate the DOA estimation performance with varying RIS element counts, maintaining $N=32$ measurements. As illustrated in Fig.~\ref{MN}(a), the performance enhances with an increased number of elements. Additionally, Fig.~\ref{MN}(b) presents the performance across different numbers of measurements, indicating improved estimation accuracy with a higher measurement count. Considering both the computational complexity of the RIS-ADMM method and the estimation performance, we recommend employing $M=64$ elements and $N=32$ measurements in the passive sensing system.

	\begin{table}
		\renewcommand{\arraystretch}{0.9}
		\caption{Computational Time (s)}
		\label{table2}
		\centering
		\begin{tabular}{cccccc}
			\hline
			SDP & FFT & OMP & $\ell_1$ norm & \textbf{RIS-ADMM}\\
			$2.378$ & $0.035$ & $0.020$ &$1.266$ & $\boldsymbol{0.202}$ \\
			\hline
		\end{tabular}
	\end{table}

\begin{figure}
		\centering
		\includegraphics[width=1.6in]{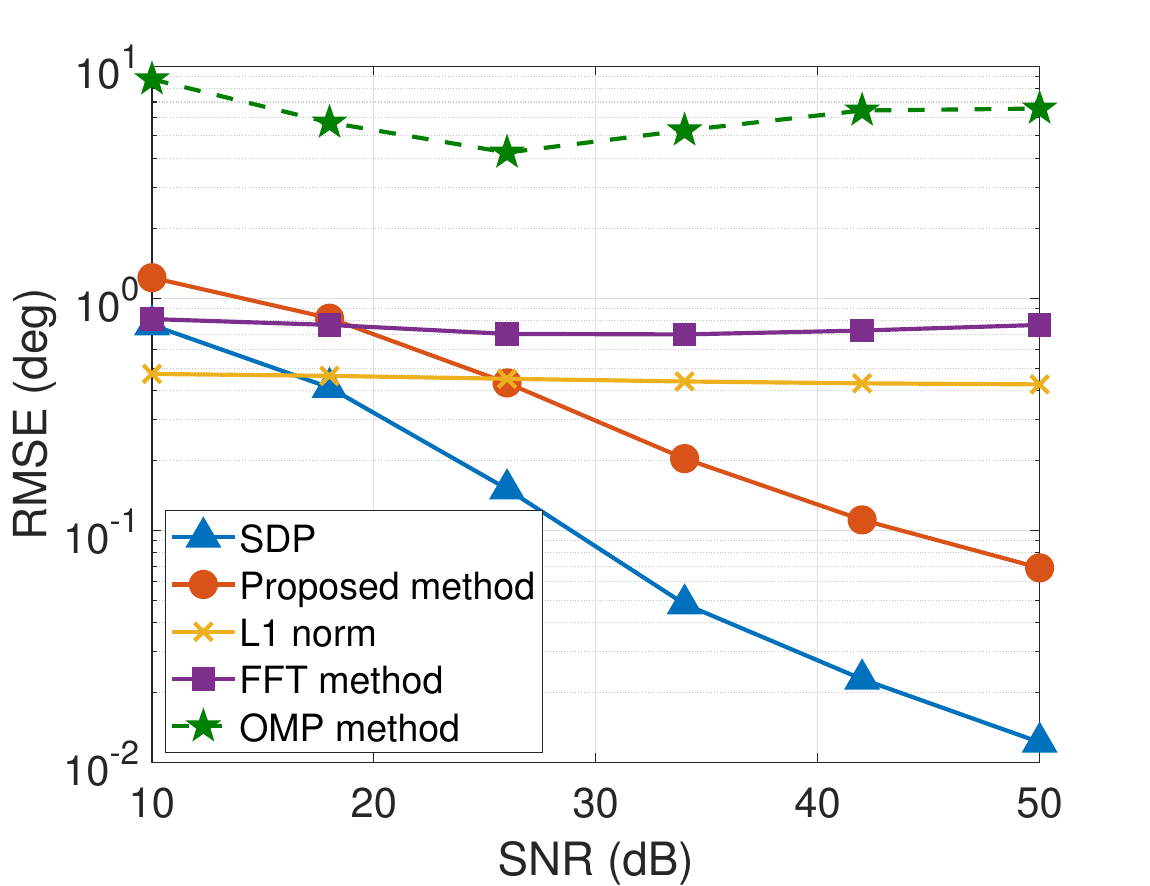}
		\caption{The estimated performance compared with other methods.}
		\label{comp}
	\end{figure}

	In this study, we conduct a comparative analysis of the proposed method with established approaches in terms of DOA estimation performance and computational efficiency. The comparative results are illustrated in Fig.~\ref{comp}, where the following $4$ methods are compared: 1) \emph{FFT method}: In this method, the interference signal is estimated first, and then the spatial spectrum is estimated by the traditional fast Fourier transformation (FFT); 2) \emph{L1 norm method}: In this method, an optimization problem is formulated by a $\ell_1$ norm minimization instead of the atomic norm, and the optimization problem is $
	\min_{\boldsymbol{x},q}\, \|\boldsymbol{r}-\boldsymbol{Gx}-\boldsymbol{c}q\|^2_2+\rho \|\boldsymbol{x}\|_{1}$;  3) \emph{OMP method}: The interference signal is also estimated first, and the orthogonal matching pursuit (OMP) method~\cite{wang2012generalized} is used to estimate the sparse signal; 4) \emph{SDP method}: The problem in (\ref{eq6}) is solved by the SDP optimization method~\cite{6576276}, and we use the CVX toolbox to solve this problem. 
	
	The proposed RIS-ADMM method demonstrates superior performance at higher SNR, effectively mitigating off-grid errors. Notably, its estimation accuracy improves with increasing SNR. However, employing only $10^2$ iterations results in a performance disparity between the proposed and SDP methods. Additionally, Table~\ref{table2} details the computational time, indicating that the proposed method's computation time is less than one-tenth that of the SDP method, and it also outperforms the $\ell_1$ norm minimization method in terms of computational complexity. Consequently, the proposed RIS-ADMM method stands out for its efficient DOA estimation, lower computational demand, and the feasibility of realizing closed-form expressions through iterative algorithms.

	\section{Conclusions}\label{sec5}
	This paper has addressed the passive sensing methodology utilizing RIS, alongside the removal of interference signals originating from AP through the formulation of an atomic norm-based optimization problem. To enhance efficiency, we introduced the RIS-ADMM algorithm, an innovative approach that iteratively solves the original SDP problem. The proposed method not only demonstrates superior DOA estimation performance but also significantly reduces computational complexity when compared to existing methodologies. Looking ahead, our future endeavors will concentrate on the practical hardware implementation of the proposed algorithm, aiming to bridge the gap between theoretical innovation and real-world application.
	
	\bibliographystyle{IEEEtran}
	\bibliography{IEEEabrv.bib,ref.bib} 
	
	\appendices
	\section{The Closed-Form RIS-ADMM Expressions} 
	\label{ap1}
	
	In the RIS-ADMM, the unknown parameters can be obtained from the following closed-form expressions:
	\begin{itemize}[topsep=-5pt]
		\item For $q^{i+1}$, let $ \nabla_{q^*}\mathcal{L}_{\tau}(\boldsymbol{x}^i,q,\boldsymbol{u}^i,t^i,\boldsymbol{z}^i,\boldsymbol{Y}^{i},\boldsymbol{P}^{i},\boldsymbol{w}^{i}) = 0.5\boldsymbol{c}^{\text{H}}\boldsymbol{w}^i+\tau\boldsymbol{c}^{\text{H}}(\boldsymbol{Gx}^i+\boldsymbol{c}q-\boldsymbol{z}^i)=0$
		and the closed form can be obtained as $ q^{i+1} = \frac{1}{\tau\|\boldsymbol{c}\|^{2}_2}\boldsymbol{c}^{\text{H}}(\tau\boldsymbol{z}^i-0.5\boldsymbol{w}^i-\tau\boldsymbol{Gx}^i)$. 
		
		\item For $\boldsymbol{x}^{i+1}$, let $\nabla_{\boldsymbol{x}^*}\mathcal{L}_{\tau}(\boldsymbol{x}^i,q^{i+1},\boldsymbol{u}^i,t^i,\boldsymbol{z}^i,\boldsymbol{Y}^{i},\boldsymbol{P}^{i},\boldsymbol{w}^{i}) =  \nabla_{\boldsymbol{x}^*}\Bigg\langle \boldsymbol{P}^i,\boldsymbol{Y}^i-\begin{pmatrix}
			\operatorname{Toep}(\boldsymbol{u}^i)&\boldsymbol{x}\\
			\boldsymbol{x}^{\text{H}} & t^i
		\end{pmatrix}\Bigg\rangle +\nabla_{\boldsymbol{x}^*}\langle \boldsymbol{Gx}+\boldsymbol{c}q^{i+1}-\boldsymbol{z}^i,\boldsymbol{w}^i\rangle + \nabla_{\boldsymbol{x}^*} \tau \left\|\boldsymbol{Y}^i-\begin{pmatrix}
			\operatorname{Toep}(\boldsymbol{u}^i)&\boldsymbol{x}\\
			\boldsymbol{x}^{\text{H}} & t^i
		\end{pmatrix}\right\|^2_{\text{F}} +\nabla_{\boldsymbol{x}^*}\tau\|\boldsymbol{Gx}+\boldsymbol{c}q^{i+1}-\boldsymbol{z}^i\|^2_2 = 0.5\boldsymbol{G}^{\text{H}}\boldsymbol{w}^i+\tau\boldsymbol{G}^{\text{H}}(\boldsymbol{Gx}+\boldsymbol{c}q^{i+1}-\boldsymbol{z}^i)  +2\tau(\boldsymbol{x}-\boldsymbol{y}^i_2)-\boldsymbol{p}^i_2=0$, 
		and $
		\boldsymbol{Y}^i \triangleq \begin{pmatrix}
			\boldsymbol{Y}^i_1 & \boldsymbol{y}^i_2\\
			\boldsymbol{y}^{i,\text{H}}_2 & y^i_3
		\end{pmatrix}$, $\boldsymbol{P}^i \triangleq \begin{pmatrix}
			\boldsymbol{P}^i_1 & \boldsymbol{p}^i_2\\
			\boldsymbol{p}^{i,\text{H}}_2 & p^i_3
		\end{pmatrix}$,  
		and the closed form can be obtained as $
		\boldsymbol{x}^{i+1}=(\tau\boldsymbol{G}^{\text{H}}\boldsymbol{G}+2\tau\boldsymbol{I})^{-1}\big[2\tau\boldsymbol{y}^i_2+\boldsymbol{p}^i_2 -\boldsymbol{G}^{\text{H}}(0.5\boldsymbol{w}^i+\tau q^{i+1} \boldsymbol{c}-\tau \boldsymbol{z}^i)\big]$. 
		\item For $\boldsymbol{u}^{i+1}$, let $\nabla_{\boldsymbol{u}^*}\mathcal{L}_{\tau}(\boldsymbol{x}^{i+1},q^{i+1},\boldsymbol{u},t^i,\boldsymbol{z}^i,\boldsymbol{Y}^{i},\boldsymbol{P}^{i},\boldsymbol{w}^{i}) =\nabla_{\boldsymbol{u}^*}\Bigg\langle \boldsymbol{P}^{i},\boldsymbol{Y}^{i}-\begin{pmatrix}
			\operatorname{Toep}(\boldsymbol{u})&\boldsymbol{x}^{i+1}\\
			\boldsymbol{x}^{i+1,\text{H}} & t^i
		\end{pmatrix}\Bigg\rangle +\nabla_{\boldsymbol{u}^*}\tau \left\|\boldsymbol{Y}^{i}-\begin{pmatrix}
			\operatorname{Toep}(\boldsymbol{u})&\boldsymbol{x}^{i+1}\\
			\boldsymbol{x}^{i+1,\text{H}} & t^{i}
		\end{pmatrix}\right\|^2_{\text{F}} +\nabla_{\boldsymbol{u}^*}\frac{\rho}{2}(u_0+t^{i})=0$, 
		and the $m$-th entry of $\boldsymbol{u}^{i+1}$ is updated as $
		u_m^{i+1} =\frac{1}{2\tau (M-m)}\Big[\operatorname{Tr}(\boldsymbol{P}^i_1,m)+2\tau\operatorname{Tr}(\boldsymbol{Y}^i_1,m -\rho e_m-2e_m\tau\operatorname{Tr}(\boldsymbol{Y}^i_1,m)\Big]$, where $e_m=\begin{cases}
			1 & m=0\\
			0 & m\neq 0
		\end{cases}$, and $\operatorname{Tr}(\boldsymbol{A},m)=\sum_{n} A_{n+m,n}$.
		
		\item For $t^{i+1}$, let $\nabla_{t}\mathcal{L}_{\tau}(\boldsymbol{x}^{i+1},q^{i+1},\boldsymbol{u}^{i+1},t,\boldsymbol{z}^i,\boldsymbol{Y}^{i},\boldsymbol{P}^{i},\boldsymbol{w}^{i})  = 0.5\rho +\nabla_{t} \Bigg\langle \boldsymbol{P}^{i},\boldsymbol{Y}^{i}-\begin{pmatrix}
			\operatorname{Toep}(\boldsymbol{u}^{i+1})&\boldsymbol{x}^{i+1}\\
			\boldsymbol{x}^{i+1,\text{H}} & t
		\end{pmatrix}\Bigg\rangle   +\nabla_{t}  \tau \left\|\boldsymbol{Y}^{i}-\begin{pmatrix}
			\operatorname{Toep}(\boldsymbol{u}^{i+1})&\boldsymbol{x}^{i+1}\\
			\boldsymbol{x}^{i+1,\text{H}} & t
		\end{pmatrix}\right\|^2_{\text{F}} =  0.5\rho -p^{i}_3 +2\tau(t-y^{i}_3)=0$, 
		and the closed form can be obtained as $ t^{i+1}=y^{i}_3+\frac{p^i_3}{2\tau}-\frac{\rho}{4\tau}$s.  
		
		\item For $\boldsymbol{z}^{i+1}$, let $\nabla_{\boldsymbol{z}^*}  \mathcal{L}_{\tau}(\boldsymbol{x}^{i+1},q^{i+1},\boldsymbol{u}^{i+1},t^{i+1},\boldsymbol{z},\boldsymbol{Y}^{i},\boldsymbol{P}^{i}$, $\boldsymbol{w}^{i}) = \nabla_{\boldsymbol{z}^*} \|\boldsymbol{r}-\boldsymbol{z}\|^2_2+\nabla_{\boldsymbol{z}^*}\langle \boldsymbol{Gx}^{i+1}+\boldsymbol{c}q^{i+1}-\boldsymbol{z},\boldsymbol{w}^{i}\rangle +\nabla_{\boldsymbol{z}^*}\tau\|\boldsymbol{Gx}^{i+1}+\boldsymbol{c}q^{i+1}-\boldsymbol{z}\|^2_2 = \boldsymbol{z}-\boldsymbol{r}-0.5\boldsymbol{w}^{i}-\tau(\boldsymbol{Gx}^{i+1}+\boldsymbol{c}q^{i+1})+\tau\boldsymbol{z}=0$, 
		and we can obtain $
		\boldsymbol{z}^{i+1}=\frac{1}{1+\tau}\left[\boldsymbol{r}+0.5\boldsymbol{w}^i+\tau(\boldsymbol{Gx}^{i+1}+\boldsymbol{c}q^{i+1})\right]$. 
		
		\item For $\boldsymbol{Y}^{i+1}$, we have $\boldsymbol{Y}^{i+1} = \arg\min_{\boldsymbol{Y}\succeq 0} \mathcal{L}_{\tau}(\boldsymbol{x}^{i+1},q^{i+1},\boldsymbol{u}^{i+1},t^{i+1},\boldsymbol{z}^{i+1},\boldsymbol{Y}, \boldsymbol{P}^{i}$, $\boldsymbol{w}^{i})  =  \arg\min_{\boldsymbol{Y}\succeq 0}  \Bigg\langle \boldsymbol{P}^i,\boldsymbol{Y}-\begin{pmatrix}
			\operatorname{Toep}(\boldsymbol{u}^{i+1})&\boldsymbol{x}^{i+1}\\
			\boldsymbol{x}^{i+1,\text{H}} & t
		\end{pmatrix}\Bigg\rangle + \tau \left\|\boldsymbol{Y}-\begin{pmatrix}
			\operatorname{Toep}(\boldsymbol{u}^{i+1})&\boldsymbol{x}^{i+1}\\
			\boldsymbol{x}^{i+1,\text{H}} & t^{i+1}
		\end{pmatrix}\right\|^2_{\text{F}}  = \arg\min_{\boldsymbol{Y}\succeq 0} \left\|\boldsymbol{Y}-\begin{pmatrix}
			\operatorname{Toep}(\boldsymbol{u}^{i+1})&\boldsymbol{x}^{i+1}\\
			\boldsymbol{x}^{i+1,\text{H}} & t^{i+1}
		\end{pmatrix}+\frac{1}{2\tau}\boldsymbol{P}^i\right\|^2_{\text{F}}$. 
		With the eigenvalue decomposition, we have $
			\begin{pmatrix}
				\operatorname{Toep}(\boldsymbol{u}^{i+1})&\boldsymbol{x}^{i+1}\\
				\boldsymbol{x}^{i+1,\text{H}} & t^{i+1}
			\end{pmatrix}-\frac{1}{2\tau}\boldsymbol{P}^i=\boldsymbol{U\Lambda U}^{\text{H}}$,
		and the optimal semidefinite matrix $\boldsymbol{Y}^{i+1}$ can be obtained by $
		\boldsymbol{Y}^{i+1}=\boldsymbol{U}\boldsymbol{\Lambda}'\boldsymbol{U}^{\text{H}}$, 
		where the diagonal matrix $\boldsymbol{\Lambda}'$ by setting the negative entries of $\boldsymbol{\Lambda}$ being $0$.
		
		\item For $\boldsymbol{P}^{i+1}$, we have $\nabla_{\boldsymbol{P}^*}\mathcal{L}_{\tau}(\boldsymbol{x}^{i+1},q^{i+1},\boldsymbol{u}^{i+1},t^{i+1},\boldsymbol{z}^{i+1}$, $\boldsymbol{Y}^{i+1},\boldsymbol{P},\boldsymbol{w}^{i}) = \nabla_{\boldsymbol{P}^*} \Bigg\langle \boldsymbol{P},\boldsymbol{Y}^{i+1}-\begin{pmatrix}
			\operatorname{Toep}(\boldsymbol{u})&\boldsymbol{x}^{i+1}\\
			\boldsymbol{x}^{i+1,\text{H}} & t^{i+1}
		\end{pmatrix}\Bigg\rangle  = \boldsymbol{Y}^{i+1}-\begin{pmatrix}
			\operatorname{Toep}(\boldsymbol{u}^{i+1})&\boldsymbol{x}^{i+1}\\
			\boldsymbol{x}^{i+1,\text{H}} & t^{i+1}
		\end{pmatrix}$, 
		so we can update $\boldsymbol{P}^{i+1}$ as $
		\boldsymbol{P}^{i+1} = \boldsymbol{P}^{i}+\rho\left[\boldsymbol{Y}^{i+1}-\begin{pmatrix}
			\operatorname{Toep}(\boldsymbol{u}^{i+1})&\boldsymbol{x}^{i+1}\\
			\boldsymbol{x}^{i+1,\text{H}} & t^{i+1}
		\end{pmatrix}\right]$. 
		
		\item For $\boldsymbol{w}^{i+1}$, we have $\nabla_{\boldsymbol{w}^*}\mathcal{L}_{\tau}(\boldsymbol{x}^{i+1},q^{i+1},\boldsymbol{u}^{i+1},t^{i+1},\boldsymbol{z}^{i+1}$, $\boldsymbol{Y}^{i+1},\boldsymbol{P},\boldsymbol{w}^{i})  = \nabla_{\boldsymbol{w}^*}\langle \boldsymbol{Gx}^{i+1}+\boldsymbol{c}q^{i+1}-\boldsymbol{z}^{i+1},\boldsymbol{w}\rangle  = 0.5(\boldsymbol{Gx}^{i+1}+\boldsymbol{c}q^{i+1}-\boldsymbol{z})^{i+1}$, 
		so we can update $\boldsymbol{w}^{i+1}$ as $\boldsymbol{w}^{i+1} = \boldsymbol{w}^{i}+0.5\rho(\boldsymbol{Gx}^{i+1}+\boldsymbol{c}q^{i+1}-\boldsymbol{z}^{i+1})$. 
	\end{itemize}

\end{document}